\pdfoutput=1
\newcommand*{\ATLASLATEXPATH}{latex/}
\documentclass[cernpreprint, atlasdraft=false, texlive=2016, UKenglish]{\ATLASLATEXPATH atlasdoc}

\usepackage{\ATLASLATEXPATH atlaspackage}
\usepackage{amsmath}
\usepackage{siunitx}
\usepackage{graphicx}
\usepackage{pdflscape}
\usepackage{\ATLASLATEXPATH atlasbiblatex}

\usepackage{\ATLASLATEXPATH atlascontribute}

\usepackage{\ATLASLATEXPATH atlasphysics}

\graphicspath{{logos/}{figures/}}

\usepackage{VLQ_Wt_2017_paper-defs}

\AtlasTitle{Search for pair production of heavy vector-like quarks decaying into high-$p_{\mathrm{T}}$ $W$ bosons and top quarks in the lepton-plus-jets final state in $pp$ collisions at $\sqrt{s}$=13~\TeV{} with the ATLAS detector}

\AtlasJournal{JHEP}

\AtlasAbstract{ %
A search is presented for the pair production of heavy vector-like $B$ quarks, primarily targeting $B$ quark decays into a $W$ boson and a top quark.
The search is based on $36.1$~fb$^{-1}$ of $pp$ collisions at $\sqrt{s}$ = 13~\TeV\ recorded in 2015 and 2016 with the ATLAS
detector at the CERN Large Hadron Collider.
Data are analysed in the lepton-plus-jets final state, characterised by a high-transverse-momentum isolated electron or muon, large missing transverse momentum, and multiple jets, of which at least one is $b$-tagged.
No significant deviation from the Standard Model expectation is observed. 
The 95\% confidence level lower limit on the $B$ mass is 1350~\GeV{} assuming a 100\% branching ratio to $Wt$. In the SU(2) singlet scenario, the lower mass limit is 1170~\GeV{}.
This search is also sensitive to a heavy vector-like $B$ quark decaying into other final states ($Zb$ and $Hb$) and thus mass limits on $B$ production are set as a function of the decay branching ratios.
The 100\% branching ratio limits are found to be also applicable to heavy vector-like  $X$ production, with charge $+$5/3, that decay into $Wt$.
}

\author{The ATLAS Collaboration}

\AtlasRefCode{EXOT-2017-34}

\PreprintIdNumber{CERN-EP-2018-088}

\AtlasDate{\today}

\hypersetup{pdftitle={ATLAS document},pdfauthor={The ATLAS Collaboration}}

\begin{document}

\maketitle

\tableofcontents

\section{Introduction}
\label{sec:intro}
The discovery of the Higgs boson by the ATLAS and CMS collaborations
is a major milestone in high-energy
physics~\cite{HIGG-2012-27,CMS-HIG-12-028}. However, the underlying
nature of electroweak symmetry breaking remains unknown. Naturalness
arguments~\cite{naturalness} require that, to avoid fine-tuning, quadratic divergences arising from radiative
corrections to the Higgs boson mass are cancelled out by one or more new particles.
Several such mechanisms have been proposed in theories beyond the Standard Model (BSM). In
supersymmetry, the cancellation comes from assigning superpartners to
the Standard Model (SM) bosons and fermions. Alternatively, Little
Higgs~\cite{ArkaniHamed:2002qy,Schmaltz:2005ky} and Composite
Higgs~\cite{Kaplan:1983sm,Agashe:2004rs} models introduce a
spontaneously broken global symmetry, with the Higgs boson emerging as
a pseudo Nambu--Goldstone boson~\cite{Hill:2002ap}.  These latter
models predict the existence of vector-like quarks (VLQs), defined as
colour-triplet spin-1/2 fermions whose left- and right-handed chiral
components have the same transformation properties under the
weak-isospin SU(2) gauge
group~\cite{delAguila:1982fs,AguilarSaavedra:2009es}.
Depending on the model, vector-like quarks are produced in SU(2) singlets, doublets, or triplets of flavours $T$, $B$, $Y$ or $X$, in which the first two have the same charge as the SM top quark and $b$-quark while the vector-like $Y$ and $X$ quarks have charge\footnote{All electric charges are quoted in units of $e$.} $-$4/3 and $+$5/3, respectively.
In addition, in these models, VLQs are expected to couple preferentially to third-generation
quarks~\cite{delAguila:1982fs,Aguilar-Saavedra:2013wba} and can have
flavour-changing neutral-current decays in addition to the
charged-current decays characteristic of chiral quarks. As a result,
an up-type $T$ quark can decay not only into a $W$
boson and a $b$-quark, but also into a $Z$ or Higgs boson and a
top quark ($T \to Wb$, $Zt$, and $Ht$). Similarly, a down-type $B$ quark can decay into a $Z$ or Higgs boson and a
$b$-quark, in addition to decaying into a $W$ boson and a top quark
($B \to Wt$, $Zb$, and $Hb$). For each type, the sum of the three branching fractions is assumed to be 1, i.e. other decays are not considered.
Due to their charge, vector-like $Y$ quarks decay exclusively into $Wb$ while vector-like $X$ quarks decay exclusively into $Wt$.  To be consistent with the
results from precision electroweak measurements, the mass-splitting
between VLQs belonging to the same SU(2) multiplet is required to be small, but no
requirement is placed on which member of the doublet is
heavier~\cite{Aguilar-Saavedra:2013qpa}.  Cascade decays such as
$T \to WB\to WWt$ are thus assumed to be kinematically forbidden.
Decays of VLQs into final states with first- and second-generation quarks,
although not favoured, are not
excluded by precision electroweak or flavour measurements~\cite{Atre:2008iu,Atre:2011ae}.

\begin{figure}[h!]
\centering
\includegraphics[width=0.44\textwidth]{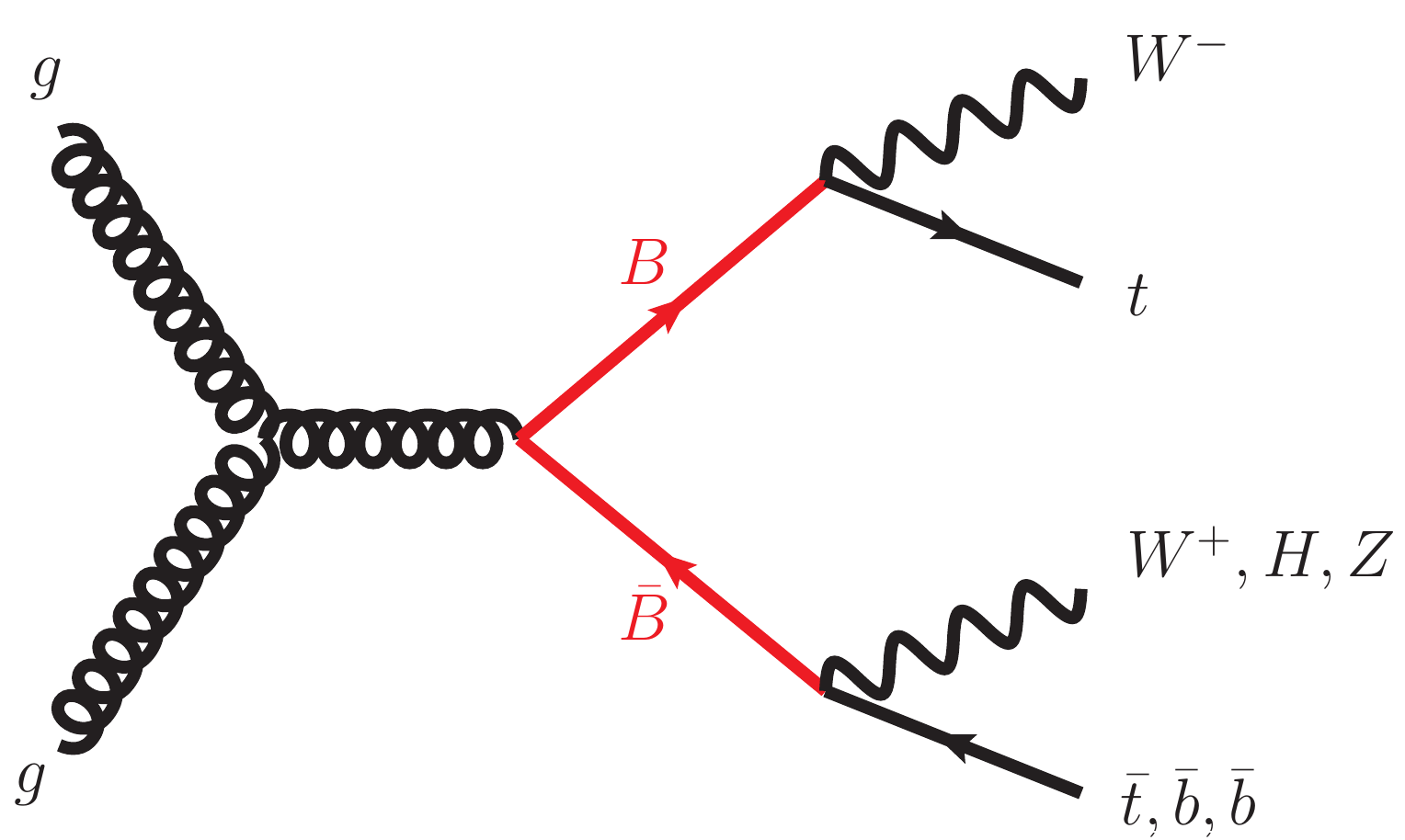}
\caption{Example of a leading-order $B\bar{B}$ production diagram in the targeted $Wt$ decay mode.}
\label{fig:feyn}
\end{figure}

This search targets the $B\to Wt$ decay mode using the $pp$ collision data collected at the
Large Hadron Collider (LHC) in 2015 and 2016 at a centre-of-mass
energy of 13~\TeV{}, although it
is also sensitive to a wide range of branching ratios to the other two
decay modes as well as to production of vector-like $X$ quarks. An example of a leading-order production diagram is shown in Figure~\ref{fig:feyn}.
Previous searches in this decay mode by the ATLAS and CMS
collaborations did not observe a significant deviation from the SM
predictions. Those searches excluded VLQ masses below 740~\GeV{} for any
combination of branching ratios and below 1020~\GeV{} for the
assumption of \BR$(B\to Wt)=1$~\cite{PhysRevD.93.112009,Sirunyan2017}.
A recent search by the ATLAS Collaboration at $\sqrt{s}=13$~\TeV{}, primarily targeting the $T$ quark decaying into $Wb$, was also found to be sensitive to $B$ and $X$ quarks decaying into $Wt$. The results included interpretations which provide a 95\% confidence-level observed (expected) lower limit on the $B$ quark mass at 1250~(1150)~\GeV{} assuming a 100\% branching ratio to $Wt$; in the SU(2) singlet scenario, the limit is 1080~(980)~\GeV{}~\cite{Wb_13Tev}.
In this context, the event selection for this new search is optimised for high-mass \BBbar{} production with
subsequent decay into two high-\pt{} $W$ bosons and two top quarks,
where one of the four $W$ bosons decays leptonically and the others decay
hadronically.  To suppress the SM background, boosted-jet reconstruction
techniques~\cite{ATL-PHYS-PUB-2015-033,ATL-PHYS-PUB-2015-053} are used
to improve the identification of hadronically decaying high-\pt{} $W$ bosons and top quarks.
The decay products of a hadronically decaying high-momentum $W$ boson are likely to be contained within a single large-radius jet.
The two signal regions used in this search are based on the number of reconstructed large-radius jets. The first signal region aims to reconstruct the \BBbar{} system using the mass of the purely hadronically decaying $B$ candidate to discriminate between SM and VLQ
events. The second, more inclusive, signal region uses a Boosted Decision Tree (BDT) to discriminate between SM and VLQ events.

Finally, a profile likelihood fit is used to test for the
presence of a VLQ signal as a function of the $B$ quark mass and the
decay branching ratios.
The results are found to be equally applicable to either singlet or doublet weak-isospin configurations as well as to the decays of $X$ quarks.


\section{ATLAS detector}
\label{sec:detector}
%
The ATLAS detector~\cite{PERF-2007-01} at the LHC is a multipurpose particle detector with a
forward-backward symmetric cylindrical geometry that covers nearly the entire solid angle 
around the collision point. It consists of an inner detector
surrounded by a thin superconducting solenoid providing a 2~T axial magnetic
field, electromagnetic and hadronic calorimeters, and a muon spectrometer.
The inner detector covers the pseudorapidity range\footnote{
The ATLAS Collaboration uses a right-handed coordinate system with its origin at the nominal
interaction point (IP) in the centre of the detector
and the $z$-axis along the beam pipe. The $x$-axis points from the IP to the
centre of the LHC ring, and the $y$-axis points upwards. Cylindrical coordinates
$(r,\phi)$ are used in the transverse plane, $\phi$ being the azimuthal angle
around the $z$-axis. The pseudorapidity is defined in terms of the polar angle
$\theta$ as $\eta=-\ln\tan(\theta/2)$. Angular distance
is  measured in units of $\Delta R \equiv  \sqrt{(\Delta\eta)^2 + (\Delta\phi)^2} $.}
 $|\eta| < 2.5$. It consists of a silicon pixel detector, including the insertable 
B-layer installed after Run 1 of the LHC~\cite{IBL_1,IBL_2}, and a silicon microstrip detector surrounding the pixel detector, 
followed by a transition radiation straw-tube tracker.
Lead/liquid-argon sampling calorimeters provide electromagnetic
energy measurements with high granularity and a hadronic (steel/scintillator-tile)
calorimeter covers the central pseudorapidity range ($|\eta| < 1.7$).
The endcap and forward regions are instrumented with liquid-argon calorimeters for
both the electromagnetic and hadronic energy measurements up to $|\eta| = 4.9$.
The outer part of the detector consists of a muon spectrometer with high-precision tracking chambers for coverage up to $|\eta| = 2.7$, fast detectors for triggering over $|\eta| < 2.4$, and three large superconducting toroid magnets with eight coils each.
The ATLAS detector has a two-level trigger system to 
select events for offline analysis~\cite{Aaboud:2016leb}.


\section{Data and simulation}
\label{sec:Samples}
Data are only used if all ATLAS detector subsystems were operational. This search utilises a data set corresponding to $36.1$~\ifb{} of integrated luminosity from $pp$ collisions at $\sqrt{s}=13$~\TeV{} collected by the ATLAS experiment, with $3.2$~\ifb{} collected in 2015 and $32.9$~\ifb{} collected in 2016.

In all simulated events used in this search, the top quark and Higgs boson masses were set to 172.5~\GeV{} and
125~\GeV{}, respectively.
Simulated \BBbar{} signal events were generated with the leading-order (LO) generator \PROTOS~v2.2~\cite{protos} using the NNPDF2.3~\cite{pdf3} LO parton distribution function (PDF)
set and a set of tuned parameters called the A14 tune~\cite{A14} and were passed to \PYTHIAV{8.186}~\cite{Pythia81} for the underlying-event description, parton showering, and fragmentation.
The samples were generated for an SU(2) singlet $B$ VLQ, but with equal branching ratios of the $B$ quark
to each final state.
To check the dependence of the results on the weak-isospin of the VLQ, one sample was also generated using the SU(2) ($T$ $B$) doublet model including only the $B$ contributions.
The signal samples are normalised to pair-production cross-sections computed using
\textsc{Top++ v2.0}~\cite{topplusplus}, including next-to-next-to-leading-order (NNLO) quantum chromodynamics (QCD) corrections and soft-gluon resummation to
next-to-next-to-leading-logarithm (NNLL) accuracy~\cite{Cacciari:2011hy,Beneke:2011mq,Baernreuther:2012ws,Czakon:2012zr,Czakon:2012pz,Czakon:2013goa},
and using the MSTW 2008 NNLO PDF set~\cite{MSTW_nnlo}.
Their cross-sections vary from $3.38\pm0.25$~pb ($m_{B}=500$~\GeV) to $0.13\pm0.02$~fb ($m_{B}=2000$~\GeV).
Theoretical uncertainties are evaluated from variations of the factorisation and renormalisation scales, as well as from uncertainties in the PDFs and
$\alphas$. The latter two represent the largest contribution to the overall theoretical uncertainty in the signal cross-sections and are calculated
using the PDF4LHC~\cite{pdf4lhc} prescription with the MSTW 2008 68\% CL NNLO, CT10 NNLO~\cite{pdf1,ct10} and NNPDF2.3 5f FFN PDF sets~\cite{pdf3}.
Two benchmark signal scenarios are considered, along with a full scan of the branching-ratio plane.  The first benchmark corresponds to a $B$ quark that decays 100\% into $Wt$ and the second corresponds to the SU(2) singlet $B$ quark scenario, which predicts branching ratios of $\sim$50\%, $\sim$25\%, $\sim$25\% to $Wt$, $Zb$ and $Hb$, respectively~\cite{Aguilar-Saavedra:2013qpa}.

The main SM backgrounds that are studied using simulated samples are \ttbar{}, \wjets{}, \zjets{}, diboson, single top quark, and $\ttbar V$ ($V$ = $W$,$Z$).
The multi-jet background is estimated using a data-driven technique discussed in Section~\ref{sec:Backgrounds}.
The nominal \ttbar{} Monte Carlo sample was generated with \POWHEGBOXV{v2}~\cite{Powheg} interfaced with \PYTHIAV{8.2}~\cite{Pythia82}
for the parton shower and hadronisation, using the \textsc{A14} tune and the
NNPDF2.3 LO PDF set, setting the next-to-leading-order (NLO) radiation factor, $h_{\mathrm{damp}}$, to 1.5 times the mass of the top quark, $m_{\text{top}}$.
To estimate \ttbar{} modelling uncertainties, described in Section~\ref{sec:systematic_modelling}, additional samples were generated
using \POWHEGBOXV{v2} interfaced with \HERWIGV{7}~\cite{Herwig},
and \MGMCatNLOV{2.1.1}~\cite{Madgraph}
interfaced with \PYTHIAV{8.2}.  In addition, samples with \POWHEGBOXV{v2} interfaced with \PYTHIAV{8.2} were generated
with the factorisation and renormalisation scales varied by factors of 2 and 0.5, as well as with $h_{\mathrm{damp}}$ varied between $1.5\times m_{\text{top}}$ and $3\times m_{\text{top}}$.
The \ttbar{} samples are normalised to the NNLO cross-section, including NNLO QCD corrections and soft-gluon resummation to NNLL accuracy, as performed for the signal samples.

Single top quark production (called `single top' in the following) in the $s$-channel and in $Wt$ final states was also generated with \POWHEGBOXV{v2} interfaced with \PYTHIAV{6.428}~\cite{Pythia6},
while single top production in the $t$-channel was generated with \POWHEGBOXV{v1} interfaced with \PYTHIAV{6.428} for the parton shower and hadronisation.
Single-top samples were generated using the \textsc{Perugia2012} tune~\cite{Perugia} and the CT10 PDF set~\cite{ct10}.
The diagram removal method was used to remove the overlap between NLO $Wt$ production and LO \ttbar{} production~\cite{DRDS}.
The single-top cross-sections for the $t$- and $s$-channels are normalised to their NLO predictions using \textsc{Hathor v2.1}~\cite{hathor1,hathor2}, while for the $Wt$ final states the cross-section is normalised to its NLO+NNLL prediction~\cite{Kidonakis:2011wy}.
For \wjets{}, \zjets{}, and diboson ($WW$, $WZ$, $ZZ$) samples, the \SHERPAV{2.2.1} generator~\cite{Sherpa}
was used with the CT10 PDF set. The \wjets{} and \zjets{} production samples are normalised to the NNLO cross-sections~\cite{Anastasiou:2003ds,Vjets_xs,Gavin:2010az}. For diboson production,
the generator cross-sections (already at NLO) are used for the sample normalisation.
The $\ttbar V$ background is modelled using samples produced with \MGMCatNLOV{2.1.1} interfaced with \PYTHIAV{8.186}, using the A14 tune and the NNPDF2.3 LO PDF set. The \ttbar{}+$V$ samples are normalised to their
respective NLO cross-sections~\cite{Madgraph}.

All simulated events, except those from \SHERPA, use \textsc{EvtGen v1.2.0}~\cite{evtgen} for the modelling of $b$-hadron decays.
All simulated event samples for the nominal predictions were produced using the ATLAS simulation infrastructure~\cite{ATLASSIM},
using the full \GEANT4~\cite{Geant4} simulation of the ATLAS detector. The alternative \ttbar\ generator samples were processed with a fast simulation~\cite{af2} of the ATLAS detector with parameterised showers in the calorimeters.
Simulated events were then reconstructed with the same software as used for the data.
Multiple overlaid $pp$ collisions in the same or nearby bunch crossings (pile-up) were simulated at rates matching those in the data; they were
modelled as low-$\pt$ multi-jet production using the \PYTHIAV{8.186} generator and the  \textsc{A2} tune~\cite{ATLAS:2012uec}.
Additional corrections are applied to the simulated samples to correct for residual deviations of efficiencies and resolutions from those observed in the data.


\section{Analysis object selection}
\label{sec:ObjectDefinitions}
Reconstructed objects are defined by combining information from different detector subsystems.  This section outlines the criteria used to identify and select the reconstructed objects used in the analysis. Events are required to have at least one vertex candidate with at least two tracks with $\pt > 400$~\MeV. The primary vertex is taken to be the vertex candidate with the largest sum of squared transverse momenta of all associated tracks.

To reconstruct jets, three-dimensional energy clusters in the calorimeter~\cite{Aad:2016upy}, assumed to represent massless particles coming from the primary vertex, are grouped together using the anti-$k_t$ clustering algorithm~\cite{antikt1,antikt2,Cacciari:2011ma} with a radius parameter of 0.4 (1.0) for small-$R$ (large-$R$) jets.  Small-$R$ jets and large-$R$ jets are clustered independently using the same inputs.

Small-$R$ jets are calibrated using an energy- and $\eta$-dependent calibration scheme, with \textit{in situ} corrections based on data~\cite{PERF-2016-04}, and
are selected if they have $\pt > 25$~\GeV{} and $|\eta| < 2.5$. A multivariate jet vertex tagger (JVT) selectively removes small-$R$ jets below 60~\GeV{} that are identified as having originated from pile-up collisions rather than the hard scatter~\cite{jvt}. Jets containing $b$-hadrons are identified via an algorithm that uses multivariate techniques to combine information from the impact parameters of displaced tracks as well as topological
properties of secondary and tertiary decay vertices reconstructed within the jet~\cite{PERF-2012-04,ATL-PHYS-PUB-2016-012}.
A jet is considered $b$-tagged if the value for the multivariate discriminant is above
the threshold corresponding to an efficiency of 77\% for tagging a jet containing $b$-hadrons. The corresponding light-jet rejection factor is $\sim130$ and the charm-jet rejection factor is $\sim 6$, as determined in simulated \ttbar{} events~\cite{ATLAS-CONF-2018-001}. 

Large-$R$ jets are built using
the energy clusters in the calorimeter and then trimmed~\cite{Krohn2010} to mitigate the
effects of contamination from pile-up and to improve background rejection.  The jet energy and pseudorapidity are
further calibrated to account for residual detector effects using energy- and pseudorapidity-dependent calibration factors derived from simulation.
The $k_t$-based trimming algorithm reclusters the jet constituents into subjets with a more finely grained resolution with an $R$-parameter set to  $R_{\text{sub}}=0.2$. Subjets that contribute less than 5\% to the \pt{} of the large-$R$ jets are discarded.
The properties (e.g.\ transverse momentum and invariant mass) of the jet are
recalculated using only the constituents of the remaining subjets.
Trimmed large-$R$ jets are only considered if they have $\pt{}>200$~\GeV{} and $|\eta|<2.0$. No dedicated overlap-removal procedure between large-$R$ and small-$R$ jets is performed.
To identify large-$R$ jets that are likely to have originated from the hadronic decay of $W$ bosons ($W_\text{had}$),
jet substructure information is exploited using both the ratio of the energy correlation functions $D_2^{\beta=1}$~\cite{Larkoski:2014gra,Larkoski:2015kga} and the combined jet mass~\cite{ATLAS-CONF-2016-035}.
The combined jet mass is constructed using a combination of the calorimeter-derived jet mass, based on calorimeter cell cluster constituents, and the track-assisted jet mass, where the calorimeter momentum is augmented by information from the tracks associated with the large-$R$ jet.
Selected large-$R$ jets must pass both the substructure and mass requirements of the 50\%-efficient $W$-tagging working point~\cite{ATL-PHYS-PUB-2015-033}.
To reduce the
contribution from the $t\bar t$ background, $W_\text{had}$ candidates
must not overlap any $b$-tagged small-$R$ jets within $\Delta R < 0.75$.


Electrons are reconstructed from energy deposits in the electromagnetic calorimeter matched to inner detector tracks.
Electron candidates are required to satisfy likelihood-based identification criteria~\cite{ATLAS-CONF-2016-024} and must have $\pt^\text{ele} > 30$~\GeV{} and $|\eta|<2.47$.
Electron candidates in the transition region between the barrel and endcap electromagnetic calorimeters, $1.37 < |\eta| < 1.52$, are excluded.
A lepton isolation requirement is implemented by calculating the quantity
$I_R =
\sum_{\Delta R
(\mathrm{track},\text{ele})
<R_\mathrm{cut}}
\pt^\mathrm{track}$,
 where $R_\mathrm{cut}$ is the smaller of 10~\GeV/$p_{\text{T}}^\text{ele}$ and 0.2; the track associated with the lepton is excluded from the calculation. The electron must satisfy $I_R < 0.06\cdot p_{\text{T}}^\text{ele}$.
Additionally, electrons are required to have a track satisfying $|d_0|/\sigma_{d_0} < 5$ and $|z_0 \sin\theta| < 0.5$\,mm, where $d_0$ is the transverse impact parameter and $z_0$ is the $r$--$\phi$ projection of the impact point onto the $z$-axis.
An overlap-removal procedure prevents double-counting of energy between an electron and nearby jets
by removing jets if the separation between the electron and jet is within $\Delta R< 0.2$ and removing electrons if the separation is within $0.2< \Delta R< 0.4$.
Subsequently, a large-$R$ jet is removed if the separation between the electron and the large-$R$ jet is within $\Delta R = 1.0$.

Muons are reconstructed from inner detector tracks matched to muon spectrometer tracks or track segments~\cite{Aad:2016jkr}. Candidate muons are required to satisfy quality specifications based on information from the muon spectrometer and inner detector.
Furthermore, muons are required to be isolated using the same criterion that is applied to electrons and their associated tracks must satisfy $|z_0 \sin\theta| < 0.5$~mm and $|d_0|/\sigma_{d_0} < 3$.
Muons are selected if they have $\pt{}>30$~\GeV{} and $|\eta|<2.5$.
An overlap-removal procedure is also applied to muons and jets.  If a muon and a jet with at least three tracks are separated by $\Delta R < \min(0.4, 0.04+10~\GeV/\pt{}^{\mu})$ the muon is removed;  if the jet has fewer than three tracks, the jet is removed.

For a given reconstructed event, the negative vector
sum of the \pt{} of all reconstructed leptons and small-$R$ jets is defined as the missing transverse momentum ($\vec{E}_\mathrm{T}^\mathrm{miss}$)~\cite{met}.
An extra term is included to account for `soft' energy from
inner detector tracks that are not matched
to any of the selected objects but are consistent with originating from the primary vertex.


\section{Analysis strategy}
\label{sec:AnalysisStrategy}
This search targets the decay of high-mass pair-produced VLQs, $\BBbar{}$, where one $B$ quark decays into $Wt$ and the other
decays into $Wt$, $Zb$ or $Hb$.
Since a recent search by ATLAS~\cite{Wb_13Tev}, primarily targeting the $T$ quark decays into $Wb$, has been reinterpreted to exclude VLQs decaying into $Wt$ at 95\% confidence level (CL) for masses below 1250~\GeV{}, this search focuses on the decays of high-mass VLQs.
The final state consists of a high-\pt{}
charged lepton and missing transverse momentum from the decay of one of the $W$ bosons, high-momentum large-$R$ jets from hadronically decaying boosted $W$ bosons, and multiple $b$-tagged jets.
The event preselection is described in Section~\ref{sec:EventSelection} and the classification of events into two
non-overlapping signal regions follows in Section~\ref{sec:ControlRegions}.


\subsection{Event preselection}
\label{sec:EventSelection}
Events were recorded using a combination of single-electron or single-muon triggers with isolation requirements.
In 2015, the lowest \pt\ threshold was 24~\GeV{}; in 2016, it ranged from 24 to 26~\GeV{}. Additional triggers
without an isolation requirement were used to recover efficiency for leptons with $\pt > 60$~\GeV{}.
Events are required to have exactly one lepton candidate (electron or muon, $N_{\mathrm{lep}}$) that must be geometrically matched to the triggering lepton.
Signal events are expected to have a high jet multiplicity ($N_{\mathrm{jets}}$), since they include up to two $b$-jets ($N_{b\text{-jets}}$) as well as jets from the hadronic decay of up to three $W$ bosons.
Therefore, at least four small-$R$ jets are required, of which at least one must be $b$-tagged.
At least one large-$R$ jet candidate, $N_{\mathrm{jets}}^\mathrm{large}$, with no $W$-tagging requirement applied, is required and the \etmiss{} is required to be greater than 60~\GeV{}.
Signal events are expected to have characteristic high values in $S_\mathrm{T}$, defined by the scalar sum of \met{} and the transverse momenta of the lepton and all small-$R$ jets. In this context, $S_\mathrm{T}$ is required to be greater than 1200~\GeV{}.

Assuming exactly one neutrino is present in each event, its four-momentum can be analytically determined using the missing transverse momentum vector $\vec{E}_\mathrm{T}^\mathrm{miss}$ and assuming the lepton--neutrino system has an invariant mass equal to that of the $W$ boson. Nearly half
of the events are found to produce two complex solutions. When complex solutions are obtained, a real solution is
determined by minimising a $\chi^2$ parameter based on the difference between the mass of the lepton--neutrino system and the nominal value of the $W$
boson mass. In the case of two real solutions, the solution with the smaller absolute value of the longitudinal momentum is used.

After this selection, backgrounds with large contributions include \ttbar{}, \wjets{}, and single-top events.
Other SM processes, including diboson, \zjets{},  \ttbarV{} and multi-jet production, make a smaller but non-negligible contribution.


\subsection{Classification of event topologies}
\label{sec:ControlRegions}
Two orthogonal signal regions are defined. The reconstructed signal region (RECOSR) aims to reconstruct the \BBbar{} system, whereas the more inclusive signal region (BDTSR) uses a BDT to discriminate between SM and VLQ events. For signal models with \BR$(B\to Wt)=1$ the relative importance of both signal regions in the final combined fit is roughly equal. In contrast, for SU(2) singlet $B$ scenarios the BDTSR dominates.
A summary of the event selection requirements is given in Table~\ref{tab:selections} and the two signal regions are described in detail in Section~\ref{sec:RECOSR} and Section~\ref{sec:BDTSR}.

\begin{table}[!ht]{}
\begin{center}
    \caption{Summary of the event selection requirements of the two signal regions.}
    \label{tab:selections}
\begin{tabular}{l|c|c}
\hline
\hline
 & RECOSR & BDTSR\\
\hline
Leptons   & \multicolumn{2}{c}{$N_{\mathrm{lep}}=1$}\\\hline
Small-$R$ jets & \multicolumn{2}{c}{$N_{\mathrm{jets}}\geq4$} \\\hline
$b$-tagged jets & \multicolumn{2}{c}{$N_{b\text{-jets}}\geq1$} \\\hline
Large-$R$ jets & \multicolumn{2}{c}{$N_{\mathrm{jets}}^\mathrm{large}\geq1$} \\\hline
   & \multicolumn{2}{c}{$E_{\mathrm{T}}^\mathrm{miss}\geq60$~\GeV}\\
 & \multicolumn{2}{c}{$S_{\mathrm{T}}\geq1200$~\GeV} \\
 \cline{2-3}
    &  $N_{\mathrm{jets}}^\mathrm{large}\geq3$  &     \\
 \raisebox{1.5ex}[0pt]{Event kinematics }  &  $N_{W_{\mathrm{had}}}\geq1$  &    \\
    &  $\Delta R(\mathrm{lep}$, leading $b$-jet$)\geq1$ & \raisebox{1.5ex}[0pt]{veto events in RECOSR} \\
    & $S_{\mathrm{T}}\geq1500$~\GeV \\
\hline
\hline
\end{tabular}
  \end{center}
\end{table}

\subsubsection{RECOSR definition}
\label{sec:RECOSR}
After the event preselection described in Section~\ref{sec:EventSelection}, further requirements are applied to reduce the contamination from SM backgrounds in events with at least three reconstructed large-$R$ jets, where at least one is required to be tagged as a $W_\text{had}$.
Events are required to have $\Delta R($lep, leading $b$-jet$) \geq 1$, as the leading $b$-jet is found to be well separated from the lepton in VLQ candidates. In addition, $S_\mathrm{T}$ is required to be greater than 1500~\GeV{}.
These requirements are found to maximise the expected sensitivity to VLQ masses above 1300~\GeV{} for events with at least three reconstructed large-$R$ jets.

The expected number of events in the RECOSR for the background processes and signal hypothesis with mass $m_B$ = 1300~\GeV{} are shown in Table~\ref{Tab:FinalYields}. For a signal model with \BR$(B\to Wt)=1$, the acceptance times efficiency of the full event selection ranges from 0.2\% to 4\% for VLQ masses from $m_B$ = 500 to 1800~\GeV.  For the SU(2) singlet $B$ scenario, for which \BR$(B\to Wt)$ is approximately 50\% for this mass range, the signal acceptance ranges from 0.1\% to 2\%. In this signal region, SM processes such as diboson, \zjets{},  \ttbarV{}, and multi-jet production, make a smaller but non-negligible contribution, and are therefore collectively referred to as `Others'.

\begin{table}[t!]{}
\begin{center}
  \caption{Event yields in the two signal regions before and after the background-only fit (see \ref{Results:fitResults}). The quoted uncertainties include statistical and systematic uncertainties; for the \ttbar{} background no cross-section uncertainty is included since it is a free parameter of the fit. The contributions from dibosons, $Z$+jets, $ttV$ and multi-jet production are included in the `Others' category for the RECOSR, whereas they are counted separately within the BDTSR\@. Modelling errors on the small  $\ttbar V$ background are neglected. In the post-fit case, the uncertainties in the individual background components can be larger than the uncertainty in the sum of the backgrounds, which is constrained by data. Both signal models correspond to $m_B =$ 1300~\GeV{}.}
\label{Tab:FinalYields}
\begin{tabular}{ l S[table-format=3.1, table-figures-uncertainty=3] S[table-format=5, table-figures-uncertainty=3]  S[table-format=3.1, table-figures-uncertainty=3] S[table-format=5, table-figures-uncertainty=3]}
\hline\hline
\multicolumn{1}{l}{} & \multicolumn{2}{c}{Pre-fit} & \multicolumn{2}{c}{Post-fit}\\
\cmidrule(rl){2-3} \cmidrule(l){4-5}
\multicolumn{1}{l}{Sample} & \multicolumn{1}{c}{RECOSR} & \multicolumn{1}{c}{BDTSR} & \multicolumn{1}{c}{RECOSR} & \multicolumn{1}{c}{BDTSR}\\
\midrule
  \ttbar  &   20.2 \pm 16 & 21 200 \pm 7300 &   19.2 \pm 5.2 & 18 300 \pm 1500\\
  \wjets   & 4.5 \pm 2.7  & 4500 \pm 2500 & 3.6 \pm 2.0  & 3600 \pm 1900 \\
  Single top   &2.4 \pm 2.4   &2100  \pm 1700 &0.8 \pm 1.0   &1000  \pm 800\\
  Others   & 2.7 \pm 1.3  &  n/a  & 2.7 \pm 1.1  &  n/a \\
  Diboson   &n/a &360 \pm 200 &n/a & 340 \pm 190 \\
  $\ttbar V$   &n/a   & 351 \pm 57 &n/a   & 350 \pm 60 \\
  \zjets   &n/a  & 310 \pm 180 &n/a  & 300 \pm 170 \\
  Multi-jet   &n/a   & 2500 \pm 1300 &n/a   & 2000 \pm 850 \\
  \midrule
  Total background & 30.0 \pm 16 & 31 300 \pm 8200 & 26.3 \pm 6.3 & 25 900 \pm 400 \\
  \midrule
   Signal (\BR$(B \rightarrow Wt)=1$)   & 7.4 \pm 0.5   & 51 \pm 2 &n/a  & n/a  \\
  Signal (SU(2) singlet)   & 2.7 \pm 0.2 & 35 \pm 1 &n/a  & n/a  \\
\midrule
    Data   & 26 & 25 863 & 26 & 25 863 \\
 \hline\hline
\end{tabular}
\end{center}
\end{table}

After the event selection, the four-momenta of the hadronic and semileptonic VLQ candidates are reconstructed using the selected  large-$R$ jets and the leptonically decaying $W$ boson candidate.
The selected large-$R$ jets are proxies for the hadronically decaying $W$ bosons and top quarks. The leptonically decaying $W$ boson ($W_\text{lep}$) candidate is reconstructed from the lepton and reconstructed neutrino. The $W_\text{lep}$ is paired with a large-$R$ jet to form the semileptonically decaying VLQ candidate. Two additional large-$R$ jets are combined to form the hadronically decaying VLQ candidate. All possible large-$R$ jet permutations are tested and the pairing that minimises the absolute value of the mass difference between the semileptonically and hadronically reconstructed VLQ candidates, $|\Delta m|$, is chosen.
It should be noted that in cases where the lepton originates from the decay of a top quark, the reconstruction described above neglects the presence of the additional $b$-jet. This was found to nonetheless provide on average the best separation between signal and background.

The final discriminating variable used in the statistical analysis is $m_B^\text{had}$, the reconstructed mass of the hadronically decaying vector-like $B$ quark candidate. This is found to provide good expected signal sensitivity.
Figure~\ref{fig:TTreco:signals} (left) shows $m_B^\text{had}$ for benchmark $B$ quark signal models and the total expected background in the RECOSR after the reconstruction algorithm is applied.
The reconstructed masses for the signal are shown to peak at the
generated $B$ quark masses.
The tails arise from misreconstructed $B$ candidates.

\begin{figure}[ht!]
\centering
\includegraphics[width=0.49\textwidth]{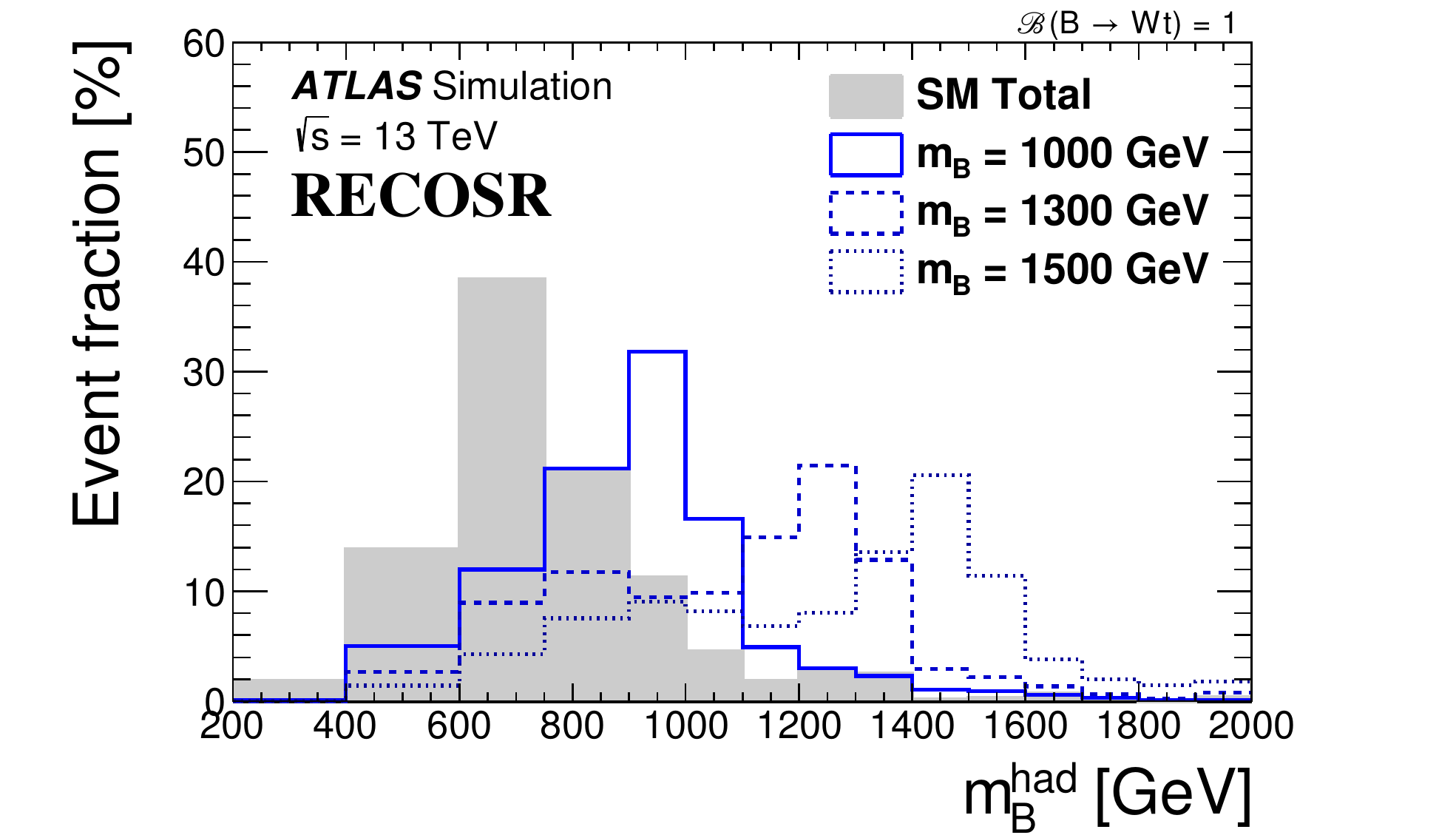}
\includegraphics[width=0.49\textwidth]{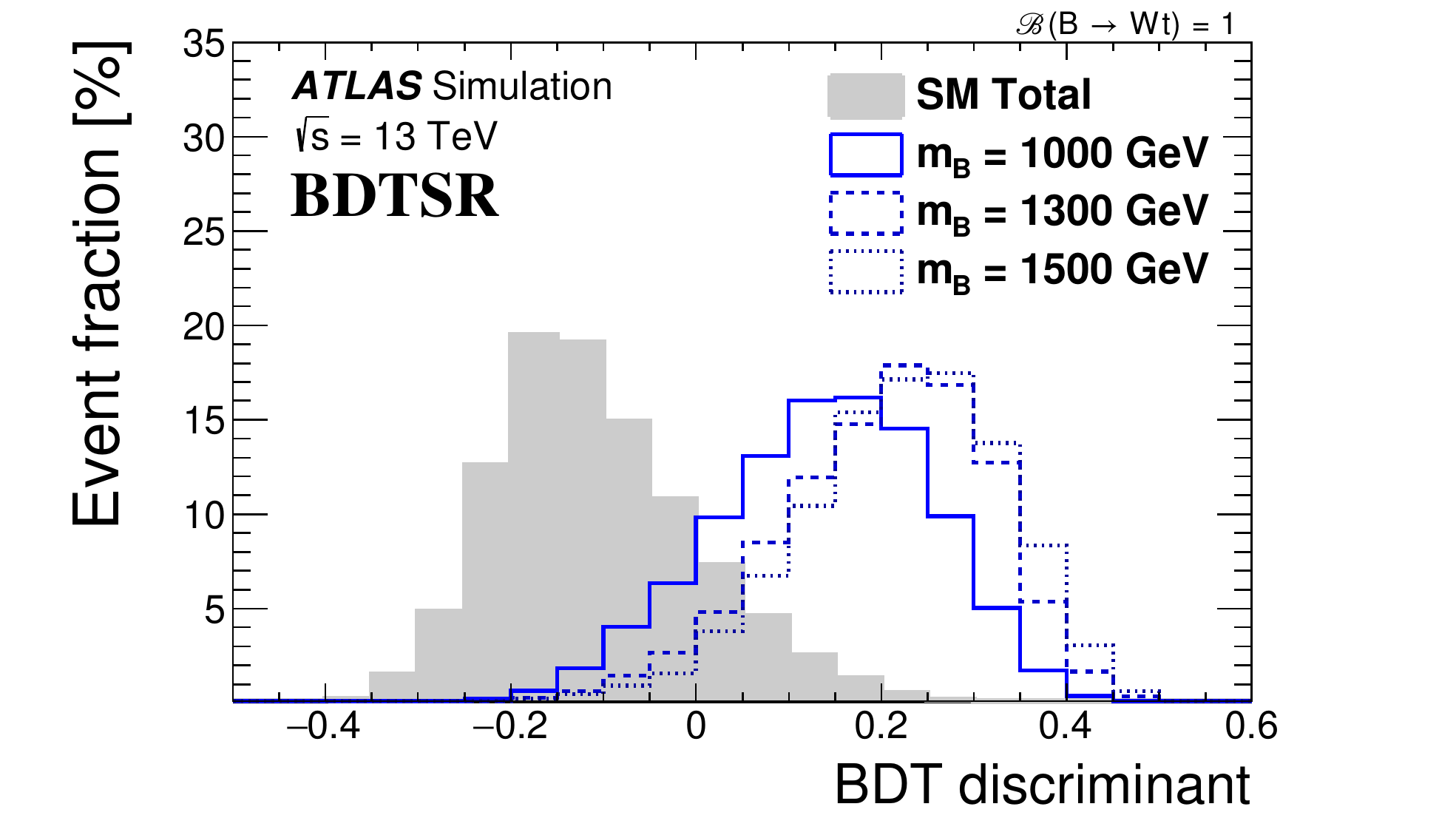}
\caption{The reconstructed hadronic $B$ quark mass in RECOSR (left) and the BDT discriminant in BDTSR (right) is shown for the total expected background and selected signal mass points for signal models with \BR$(B \rightarrow W t)= 1$. In both figures, the distributions are normalised to unity for comparison of the relative shapes at each mass point.}
\label{fig:TTreco:signals}
\end{figure}

\subsubsection{BDTSR definition}
\label{sec:BDTSR}

The BDTSR is defined by all events passing the preselection requirements (Section~\ref{sec:EventSelection}), but vetoing events contained in the RECOSR\@.
It contains events with less than three large-$R$ jets and thus it is not possible to reconstruct the full \BBbar{} system from the reconstructed objects alone. As a result, a BDT as implemented in the toolkit for multivariate data analysis with \textsc{ROOT} (\textsc{TMVA})~\cite{TMVA} is used to discriminate between potential signal and background events. For training and testing, a set of signal simulation samples assuming \BR($B \to Wt$) = 1 is used, combining signal masses ranging from 1050~\GeV{} to 1600~\GeV{}. Simulated \ttbar{} events are used as background in the training, as they are the dominant background contribution in this region.
Starting from a list of 75 variables describing the kinematics of the event, individual variables are removed through an iterative process and the performance of the BDT is evaluated, until a final set of 20 variables is selected. The procedure for removing variables is based on a combination of poor separation power, or high correlation with a variable with higher separation power, particularly if the correlation is similar between signal and background. Variables with poor agreement between data and simulation are also rejected. The 20 remaining input variables are well modelled by the simulation. The selected variables describe the global event characteristics as well as the kinematics and angular separation of the reconstructed objects. The five highest-ranked variables are:
$S_\mathrm{T}$,
the invariant mass of the highest-\pt\ large-$R$ jet,
the sphericity of the event,\footnote{Sphericity ($\mathcal{S}=\frac{2}{3}(\lambda_{2}+\lambda_{3})$) is a measure of the total momentum transverse to the sphericity axis defined by the four-momenta used for the event-shape measurement; $\lambda_{2,3}$ are the two smallest eigenvalues of the normalised momentum tensor of the small-$R$ jets, the lepton and the neutrino~\cite{sphericity2}.}
$\Delta R$ between the lepton and the sub-leading small-$R$ jet,
and
$\Delta R$ between the leading $b$-tagged jet and the leading large-$R$ jet.

The expected numbers of events in the BDTSR for the background processes and signal hypothesis with mass $m_B$ = 1300~\GeV{} are shown in Table~\ref{Tab:FinalYields}. For a signal model with \BR$(B\to Wt)=1$, the acceptance times efficiency of the full event selection ranges from 7\% to 24\% for VLQ masses from $m_B$ = 500 to 1800~\GeV.  For the SU(2) singlet $B$ scenario the signal acceptance ranges from 4\% to 16\%.

The final discriminating variable used in the statistical analysis is the BDT discriminant, which is shown in Figure~\ref{fig:TTreco:signals} (right) for benchmark $B$ quark signal models and the total expected background.

\subsection{Multi-jet background estimation}
\label{sec:Backgrounds}
The multi-jet background originates from either the
misidentification of a jet or photon as a lepton candidate (fake lepton) or from the presence of a
non-prompt lepton (e.g.\ from a semileptonic $b$- or $c$-hadron decay)
that passes the isolation requirement.  The multi-jet shape,
normalisation, and related systematic uncertainties are estimated from
data using the matrix method (MM)~\cite{Aad:2010ey}.  The MM
exploits the difference in efficiency for prompt leptons to pass loose and tight
quality requirements, obtained from $W$ and
$Z$ boson decays, and non-prompt or fake lepton candidates, from
the misidentification of photons or jets.
The efficiencies, measured in dedicated control regions, are
parameterised as functions of the lepton candidate $\pt$ and $\eta$,
$\Delta \phi$ between the lepton and jets, and the $b$-tagged jet
multiplicity.

The event selection significantly reduces the contribution of the multi-jet background in the RECOSR, to the point where statistical uncertainties make the MM prediction unreliable.
To obtain a reliable prediction, the requirement on the $W$-tagged large-$R$ jet is removed. In this region the MM prediction and the small simulation-derived backgrounds (diboson, $Z$+jets and $ttV$) are studied and their distribution shapes of the final discriminant $m_B^\text{had}$ are found to be compatible.
This selection is also used to determine the ratio of the multi-jet production to the small simulation-derived backgrounds. The ratio is then assumed to be the same in the RECOSR and is used to scale those small simulation-derived backgrounds to account for the additional contribution from multi-jet backgrounds. This scaling was found to be stable under small changes to the definition of the looser selection. In the RECOSR region, the contribution from the multi-jet background to the total background is around 1.3\%.
In the BDTSR, in contrast, the contribution of the multi-jet background is taken directly from the MM prediction


\section{Systematic uncertainties}
\label{sec:Systematics}
The systematic uncertainties are broken down into four broad categories: luminosity and cross-section uncertainties,
detector-related experimental uncertainties, uncertainties in data-driven background estimations, and modelling uncertainties in simulated background
processes. Each source of uncertainty is treated as
a nuisance parameter in the fit of the hadronic $B$ mass and BDT disciminant distributions, and shape effects are taken into account where relevant.
Due to the tight selection criteria applied, the systematic uncertainties only mildly degrade the sensitivity of the search.

\subsection{Luminosity and normalisation uncertainties}
\label{sec:systematic_Luminosity}

The uncertainty in the combined 2015+2016 integrated luminosity is 2.1\%. It is derived, following a methodology similar to that detailed in Ref.~\cite{lumi8tev}, from a preliminary calibration of the luminosity scale using $x$--$y$ beam-separation scans performed in August 2015 and May 2016.
This systematic uncertainty is applied to all backgrounds and signals that are estimated using simulated
Monte Carlo events, which are normalised to the measured integrated luminosity.

Theoretical cross-section uncertainties are applied to the relevant simulated samples.
The uncertainties for $W$/$Z$+jets and diboson production are 50\%~\cite{ATL-PHYS-PUB-2017-006,ATL-PHYS-PUB-2016-002}.
The uncertainty in the $W$+jets normalisation has a pre-fit impact\footnote{The pre-fit effect on the signal strength parameter $\mu$ is calculated by fixing the corresponding nuisance parameter at $\theta\pm \sigma_{\theta}$, where $\theta$ is the initial value of the nuisance parameter and $\sigma_{\theta}$ is its pre-fit uncertainty, and performing the fit again. The difference between the default and the modified value of $\mu$, $\Delta\mu$, represents the effect on $\mu$ of this particular uncertainty (see Section~\ref{Results:StatisticalMethod} for further details).} of 8\% on the measured signal strength for a $B$ quark mass of 1.3~\TeV{} (\BR$(B \to Wt) = 1$).
This same signal mass and branching ratio is used to quantify the impact of the uncertainties for the remainder of this section.
For single top production, the uncertainties are taken as 7\%~\cite{hathor1,hathor2}.
The normalisation of \ttbar{} is determined from the fit.
For the data-driven multi-jet estimation, an uncertainty of 100\% is assigned to the normalisation in the RECOSR, corresponding to the maximum range obtained by varying the requirements on $S_\mathrm{T}$ and $\Delta R($lep, leading $b$-jet$)$ when obtaining the multi-jet contribution from the `Others' background.
The corresponding uncertainty in the BDTSR is 50\% and evaluated by comparing the data with simulation in a region enriched in multi-jet events.

\subsection{Detector-related uncertainties}
\label{sec:systematic_detector}
The dominant sources of detector-related uncertainties in the signal and background yields relate to the small-$R$ and large-$R$ jet energy scales and resolutions.
The small-$R$ and large-$R$ jet energy scales and their uncertainties are derived by combining information from test-beam data, LHC collision data and simulation~\cite{ATL-PHYS-PUB-2015-015}.
In addition to energy scale and resolution uncertainties, there are also uncertainties in the large-$R$ jet mass and substructure scales and resolutions.
These are evaluated in a similar way to the jet energy scale and resolution uncertainties and are propagated to the $W$-tagging efficiencies.
The uncertainty in the large-$R$ jet kinematics due to differences between data and simulation seen in the large-$R$ jet calibration analysis has the largest pre-fit impact on the measured signal strength, at $\sim$12\%.

Other detector-related uncertainties come from lepton trigger efficiencies, identification efficiencies, energy scales and resolutions, the \met{} reconstruction, the $b$-tagging efficiency, and the JVT requirement. These have negligible pre-fit impact on the measured signal strength ($<$1\%).

\subsection{Generator modelling uncertainties}
\label{sec:systematic_modelling}

Modelling uncertainties are estimated for the dominant \ttbar{} and single-top backgrounds.
The modelling uncertainties are estimated by comparing simulated samples generated with different configurations, described in Section~\ref{sec:Samples}.
The effects of extra initial- and final-state gluon radiation are estimated by comparing simulated samples generated with enhanced or reduced initial-state radiation, changes to the $h_\text{damp}$ parameter, and different values of the radiation parameters. This uncertainty has a 30\% and 20\% normalisation impact on \ttbar{} in the RECOSR and BDTSR, respectively, resulting in a pre-fit impact of $\sim$3\% on the measured signal strength\footnote{The impact on the \ttbar{} normalisation is provided for illustration purposes only, as the overall \ttbar{} normalisation is a free parameter of the fit.}.
The uncertainty in the fragmentation, hadronisation and underlying-event modelling is estimated by comparing two different parton shower models, \PYTHIA\ and \HERWIGV{7}, while keeping the same hard-scatter matrix-element generator. This causes a 55\% and 5\% shift in the normalisation of \ttbar{} in the RECOSR and BDTSR, respectively, resulting in a pre-fit impact of 9\% on the measured signal strength.
The uncertainty in the hard-scatter generation is estimated by comparing events generated with two different Monte Carlo generators, \MGMCatNLO\ and \POWHEGBOX, while keeping the same parton shower model. This uncertainty has a 27\% normalisation impact on \ttbar{} in both signal regions, resulting in a pre-fit impact of $\sim$4\% on the measured signal strength.

For single top production, the dominant contribution in this analysis is from $Wt$ production and the largest uncertainty comes from the method used to remove the overlap between NLO $Wt$ production and LO \ttbar{} production. The default method of diagram removal is compared with the alternative method of diagram subtraction~\cite{DRDS}.  The full difference between the two methods is assigned as an uncertainty. This uncertainty has a 90\% and 80\% normalisation impact on single top in the RECOSR and BDTSR, respectively, resulting in a pre-fit impact of $\sim$16\% on the measured signal strength.


\section{Results}
\label{sec:Results}
\subsection{Statistical interpretation}\label{Results:StatisticalMethod}

The binned distributions of the reconstructed mass of the hadronically decaying $B$ quark candidate, $m_B^{\mathrm{had}}$, in the RECOSR, and of the BDT discriminant in the BDTSR, are used to test for the presence of a signal. Hypothesis testing is performed using a modified frequentist method as implemented in \textsc{RooStats}~\cite{RooFit,RooFitManual} and is based on a profile likelihood that takes into account the systematic uncertainties as nuisance parameters that are fitted to the data.
 A simultaneous fit is performed in the two signal regions.
The number and edges of the bins are optimised to maximise the expected vector-like $B$ quark sensitivity while ensuring the overall Monte Carlo statistical uncertainty in each bin remains below 30\%.

The statistical analysis is based on a binned likelihood function $\mathcal{L}(\mu,\theta)$ constructed as a product of Poisson probability terms over all bins considered in the search. This function depends on the signal strength parameter $\mu$, a multiplicative factor applied to the theoretical signal production cross-section, and $\theta$, a set of nuisance parameters that encode the effect of systematic uncertainties in the signal and background expectations and are implemented in the likelihood function as Gaussian constraints. Uncertainties in each bin of the fitted distributions due to the finite size of the simulated event samples are also taken into account via additional dedicated fit parameters and are propagated to $\mu$.
There are enough events in the low mass and low BDT score regions, where the signal contribution is small, to obtain a data-driven estimate of the \ttbar{} normalisation and hence the normalisation of the dominant \ttbar\ background is included as an unconstrained nuisance parameter.
Nuisance parameters representing systematic uncertainties are only included in the likelihood if either of the following conditions are met: the overall impact on the sample normalisation is larger than 1\%, or the variation induces changes
of more than 1\% between adjacent bins. This reduction of the number of nuisance parameter is done separately for the two signal regions and for each template (signal or background).
When the bin-by-bin statistical variation of a given uncertainty is significant, a smoothing algorithm is applied.

The expected number of events in a given bin depends on $\mu$ and $\theta$. The nuisance parameters $\theta$ adjust the expectations for signal and background according to the corresponding systematic uncertainties, and their values correspond to the values that best fit the data.

The test statistic $q_{\mu}$ is defined as the profile likelihood ratio,
$q_{\mu}=-2\mathrm{ln}(\mathcal{L}(\mu,\hat{\hat{\theta}}_{\mu})/ \mathcal{L}(\hat{\mu}, \hat{\theta}))$,
where $\hat{\mu}$ and $\hat{\theta}$ are the values of the parameters that maximise the likelihood function (with the constraint 0$\leq  \hat{\mu} \leq \mu$), and $\hat{\hat{\theta}}_{\mu}$ are the values of the nuisance parameters that maximise the likelihood function for a given value of $\mu$. The compatibility of the observed data with the background-only hypothesis is tested by setting $\mu=0$ in the profile likelihood ratio: $q_{0}=-2\mathrm{ln}(\mathcal{L}(0,{\hat{\hat{\theta}}}_{0})/\mathcal{L}({\hat{\mu}},{\hat{\theta}}))$.
Upper limits on the signal production cross-section for each of the signal scenarios considered are derived by using $q_{\mu}$ in the CL$_{\mathrm{s}}$ method~\cite{cls_2, cls}.
For a given signal scenario, values of the production cross-section (parameterised by $\mu$) yielding CL$_{\mathrm{s}}<0.05$, where CL$_{\mathrm{s}}$  is computed using the asymptotic approximation~\cite{cls_3}, are excluded at $\geq$ 95\% CL.

\subsection{Likelihood fit results}\label{Results:fitResults}

The expected and observed event yields in both signal regions after fitting the background-only hypothesis to data, including all uncertainties, are listed in Table~\ref{Tab:FinalYields}. The total uncertainty shown in the table is the uncertainty obtained from the full fit, and is therefore not identical to the sum in quadrature of all components, due to the correlations between the fit parameters.
The probability that the data is compatible with the background-only hypothesis is estimated by integrating the distribution of the test statistic, approximated using the asymptotic formulae~\cite{cls_3}, above the observed value of $q_{0}$. This value is computed for each signal scenario considered, defined by the assumed mass of the heavy quark and the three decay branching ratios. The lowest $p$-value is found to be $\sim$50\%, for a $B$ mass of 800~\GeV{}.
Thus no significant excess above the background expectation is found.

Individual uncertainties are generally not significantly constrained by data, except for the uncertainty associated with the single top modelling, which is constrained to be within 50\% of its initial size.

A comparison of the post-fit agreement between data and prediction for both regions is shown in Figure~\ref{FigPostFitResult}. The RECOSR shows a slight deficit of data for the $m_{B}^{\mathrm{had}}$ distribution above 800~\GeV.
Hence, the observed upper limits on the $B\bar{B}$ production cross-section are slightly stronger than the expected sensitivity.
The post-fit \ttbar{} normalisation in these regions is found to be 0.92 $\pm$ 0.30 times the Monte Carlo prediction, normalised to the NNLO+NNLL cross-section.

\begin{figure}[htb!]
\centering
\includegraphics[width=0.48\textwidth]{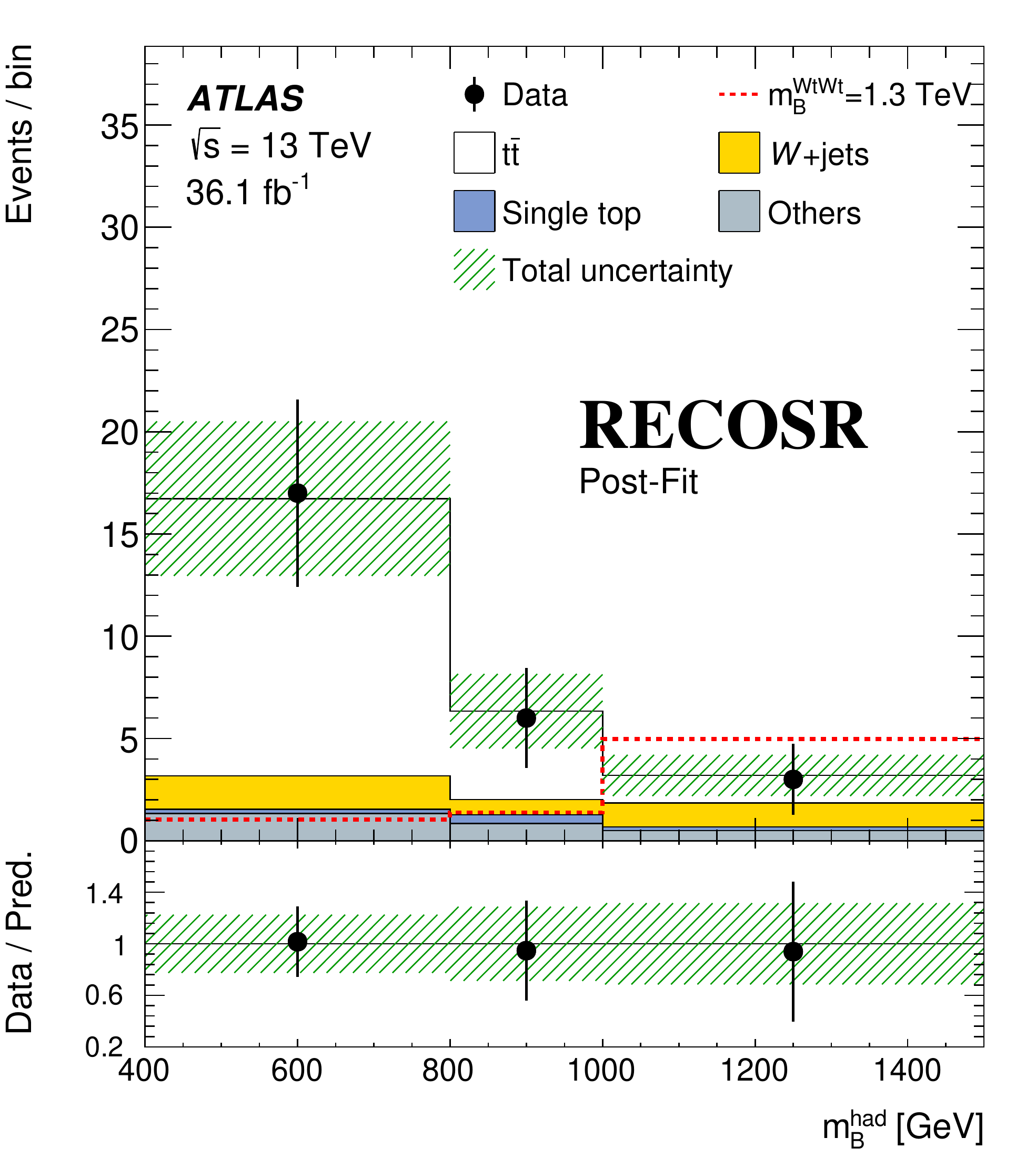}
\includegraphics[width=0.48\textwidth]{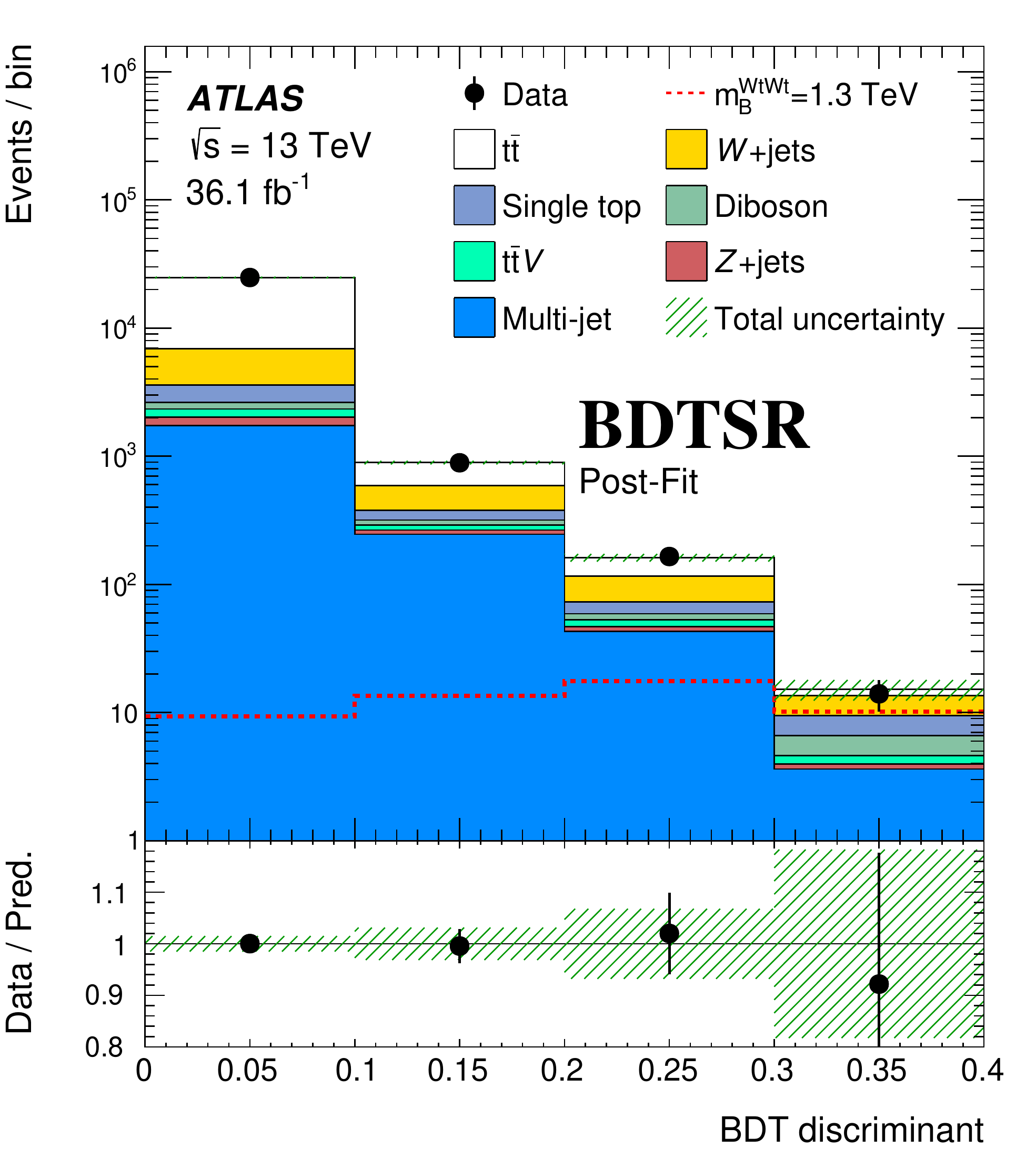}
\caption{Fit results (background-only) for the hadronic $B$ quark candidate mass distributions $m_B^{\mathrm{had}}$ (left) and the BDT discriminant in BDTSR (right). The lower panel shows the ratio of data to the fitted background yields. The band represents the total uncertainty after the maximum-likelihood fit. Events in the overflow and underflow bins are included in the last and first bin of the histograms, respectively. The expected $B\bar{B}$ signal corresponding to $m_{B} = 1300$ \GeV{} for a branching ratio of 100\% into $WtWt$ is also shown overlaid.}
\label{FigPostFitResult}
\end{figure}

\subsection{Limits on VLQ pair production}\label{Results:Limits}

Upper limits at the 95\% CL on the $B\bar{B}$ production cross-section are set for two benchmark scenarios as a function of $B$ quark
mass $m_B$ and compared with the theoretical prediction from \textsc{Top++ v2.0} (Figure~\ref{FigLimitXS}). The resulting lower limit on $m_B$ is
determined using the central value of the theoretical cross-section prediction.
These results are only valid for new particles of narrow width.
Assuming \BR($B \to Wt$) =1, the observed (expected) lower limit is $m_{B} = 1350$~\GeV{} $(1330~\GeV{})$.  For branching ratios corresponding to the SU(2) singlet $B$ scenario, the observed (expected) 95\% CL lower limit is $m_{B} = 1170$~\GeV{} $(1140~\GeV)$.  These represent a significant improvement over the reinterpreted search~\cite{Wb_13Tev}, for which the observed 95\% CL limit was 1250~\GeV\ when assuming \BR($B \to Wt$) = 1.

To check that the results do not depend on the weak-isospin of the $B$ quark
in the simulated signal events, a sample of \BBbar{} events with a
mass of 1200~\GeV{} was generated for an SU(2) ($T$ $B$) doublet $B$ quark and
compared with the nominal sample of the same mass generated for an
SU(2) singlet $B$ quark. Both the expected number of events and expected
excluded cross-section are found to be consistent between those two
samples.  Thus the limits obtained are also applicable to VLQ models with
non-zero weak-isospin.
As there is no explicit use of charge
identification, the \BR$(B \to Wt) = 1$ limits are found to be
applicable to pair-produced vector-like $X$ quarks of charge
$+5/3$ which decay exclusively into $Wt$.

Exclusion limits on $B$ quark pair production are also obtained for different values of $m_B$ and as a function of branching ratios to each of the three decays. In order to probe the complete branching-ratio plane, the signal samples are weighted by the ratios of the respective branching ratios to the original branching ratios in \PROTOS. Then, the complete analysis is repeated for each point in the branching-ratio plane.
Figure~\ref{FigLimitGrid} shows the corresponding expected and observed $B$ quark mass limits in the plane \BR($B \to Hb$) versus \BR($B \to Wt$), obtained by linear interpolation of the calculated CL$_{\mathrm{s}}$ versus $m_{B}$.

\begin{figure}[h!]
\centering
\includegraphics[width=0.7\textwidth]{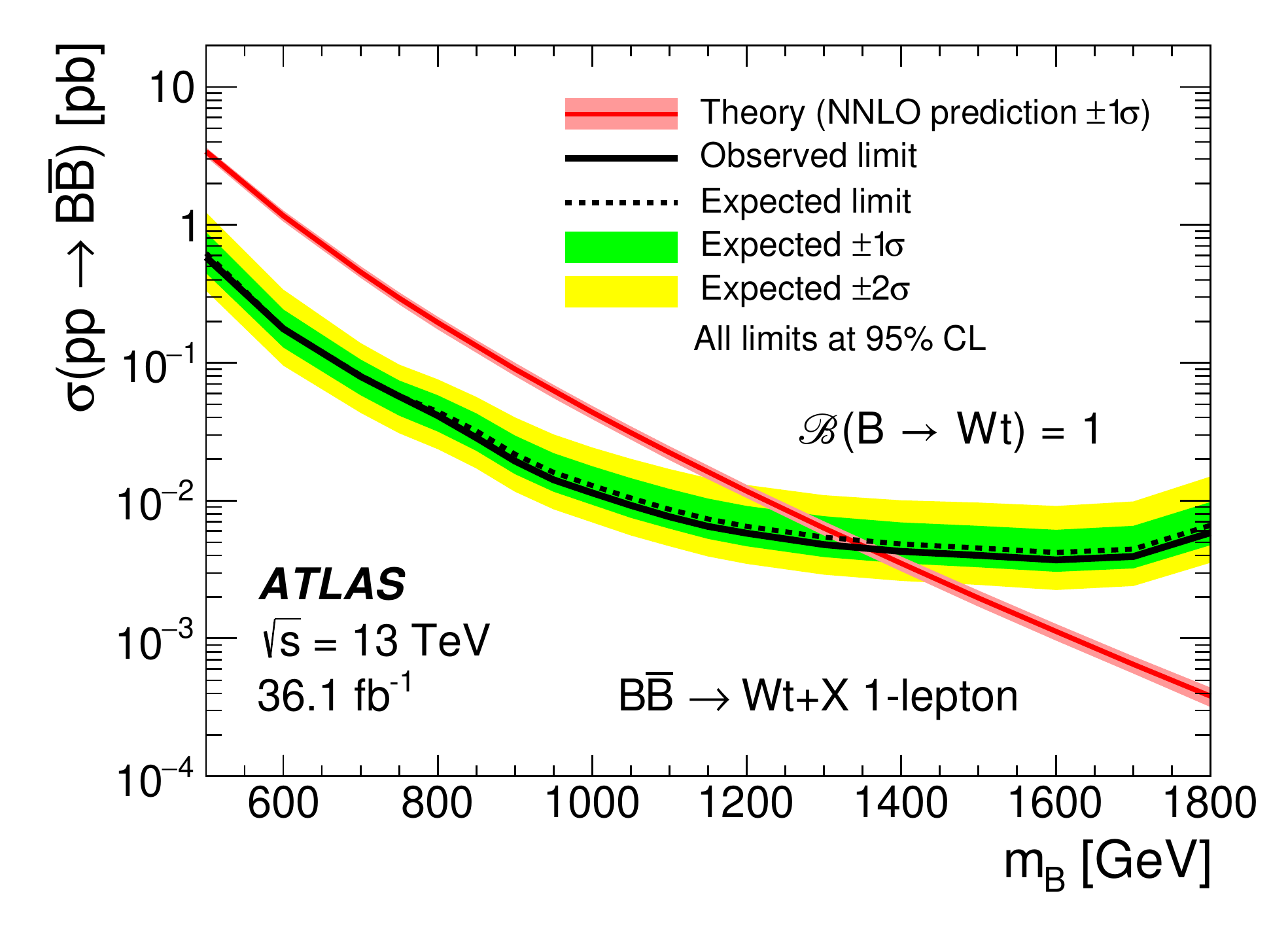}
\includegraphics[width=0.7\textwidth]{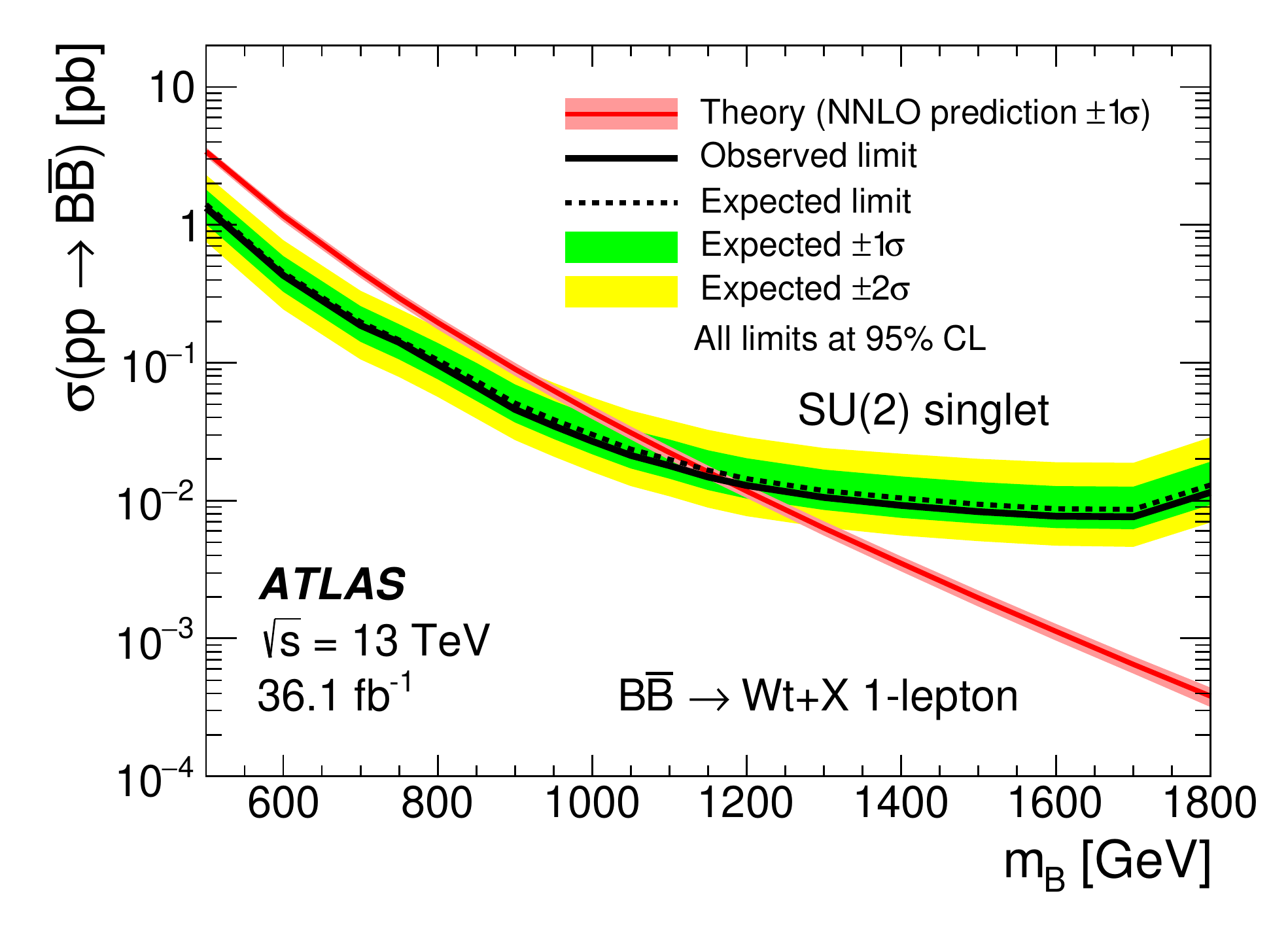}
\caption{Observed (solid line) and expected (dashed line) 95\% CL upper limits on the $B\bar{B}$ cross-section as a function of $B$ quark mass assuming \BR$(B \to Wt) = 1$ (top) and in the SU(2) singlet $B$ scenario (bottom). The surrounding shaded bands correspond to $\pm$1 and $\pm$2 standard deviations around the expected limit. The red line and band show the theoretical prediction and its $\pm$1 standard deviation uncertainty.}
\label{FigLimitXS}
\end{figure}

\begin{figure}[h!]
\centering
\includegraphics[width=0.7\textwidth]{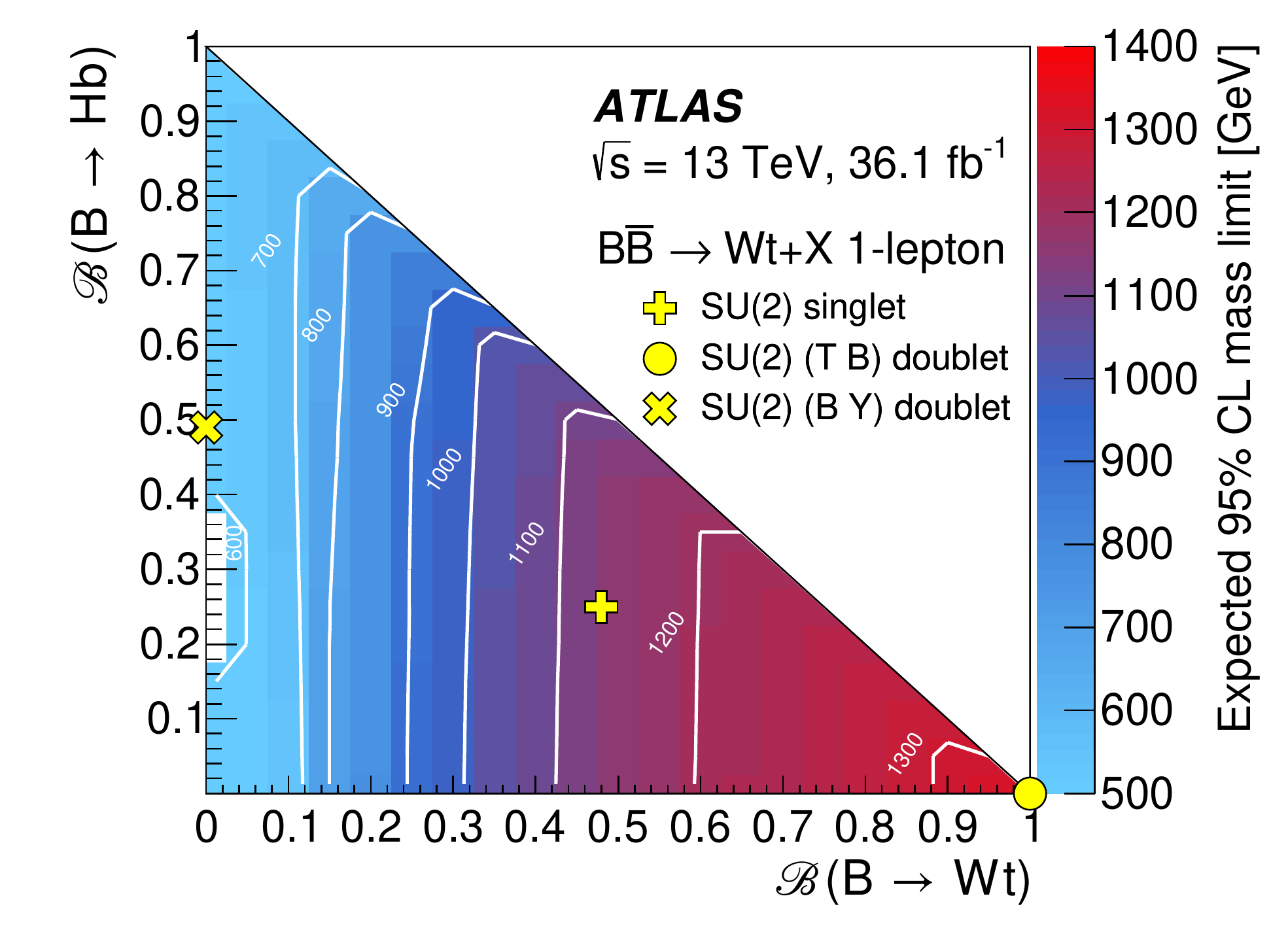}
\includegraphics[width=0.7\textwidth]{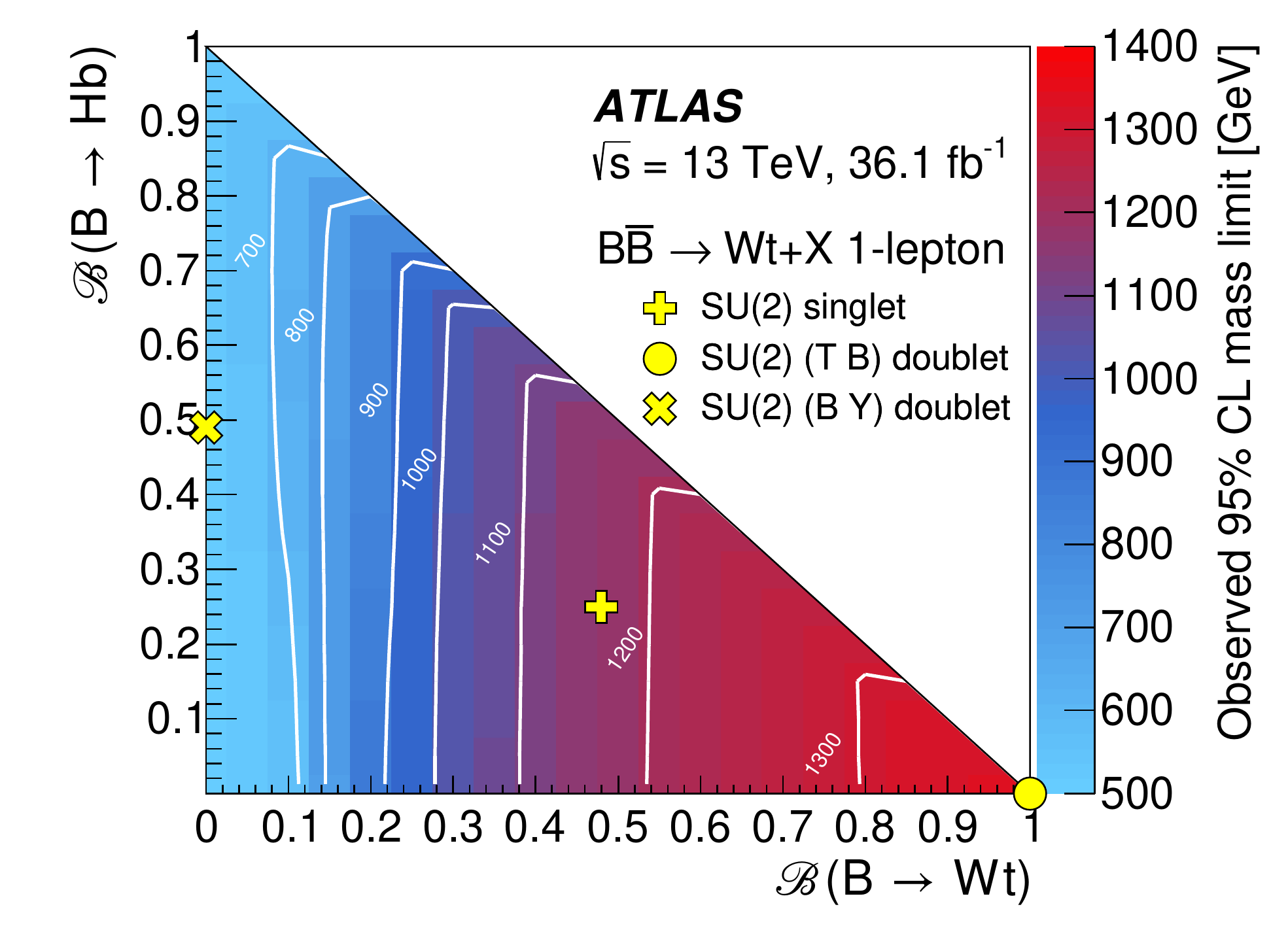}
\caption{Expected (top) and observed (bottom) 95\% CL lower limits on the mass of the $B$ quark as a function of the decay branching ratios \BR($B \to Wt$) and \BR($B \to Hb$). The white contour lines represent constant mass limits. The markers indicate the branching ratios for the SU(2) singlet and both SU(2) doublet scenarios with masses above $\sim$800~\GeV{}, where they are approximately independent of the VLQ $B$ mass.  The small white region in the upper plot is due to the limit falling below 500~\GeV{}, the lowest simulated signal mass. }
\label{FigLimitGrid}
\end{figure}


 \FloatBarrier
 \section{Conclusions}
\label{sec:conclusion}
A search for the pair production of a heavy vector-like $B$ quark, based on $pp$ collisions at $\sqrt{s}
= 13$~\TeV{} recorded in 2015 (3.2~\ifb{}) and 2016 (32.9~\ifb{})
with the ATLAS detector at the CERN Large Hadron Collider, is
presented. Data are analysed in the lepton-plus-jets final state and no significant deviation from the Standard Model
expectation is observed.  Assuming a branching ratio \BR$(B\to Wt) =
1$, the observed (expected) 95\% CL lower limit on the vector-like
quark mass is 1350~\GeV{} (1330~\GeV).  For the scenario of an SU(2)
singlet $B$ quark, the observed (expected) mass limit is 1170~\GeV{}
(1140~\GeV).  Assuming the $B$ quark can only decay into
$Wt$, $Zb$ and $Hb$, 95\% CL lower limits are derived for
various masses in the two-dimensional plane of \BR($B\to Wt$) versus
\BR($B\to Hb$). The limit for \BR$(B\to Wt) = 1$ is found to be equally applicable to VLQ $X$ quarks that decay into $Wt$. 

\section*{Acknowledgements}

We thank CERN for the very successful operation of the LHC, as well as the
support staff from our institutions without whom ATLAS could not be
operated efficiently.

We acknowledge the support of ANPCyT, Argentina; YerPhI, Armenia; ARC, Australia; BMWFW and FWF, Austria; ANAS, Azerbaijan; SSTC, Belarus; CNPq and FAPESP, Brazil; NSERC, NRC and CFI, Canada; CERN; CONICYT, Chile; CAS, MOST and NSFC, China; COLCIENCIAS, Colombia; MSMT CR, MPO CR and VSC CR, Czech Republic; DNRF and DNSRC, Denmark; IN2P3-CNRS, CEA-DRF/IRFU, France; SRNSFG, Georgia; BMBF, HGF, and MPG, Germany; GSRT, Greece; RGC, Hong Kong SAR, China; ISF, I-CORE and Benoziyo Center, Israel; INFN, Italy; MEXT and JSPS, Japan; CNRST, Morocco; NWO, Netherlands; RCN, Norway; MNiSW and NCN, Poland; FCT, Portugal; MNE/IFA, Romania; MES of Russia and NRC KI, Russian Federation; JINR; MESTD, Serbia; MSSR, Slovakia; ARRS and MIZ\v{S}, Slovenia; DST/NRF, South Africa; MINECO, Spain; SRC and Wallenberg Foundation, Sweden; SERI, SNSF and Cantons of Bern and Geneva, Switzerland; MOST, Taiwan; TAEK, Turkey; STFC, United Kingdom; DOE and NSF, United States of America. In addition, individual groups and members have received support from BCKDF, the Canada Council, CANARIE, CRC, Compute Canada, FQRNT, and the Ontario Innovation Trust, Canada; EPLANET, ERC, ERDF, FP7, Horizon 2020 and Marie Sk{\l}odowska-Curie Actions, European Union; Investissements d'Avenir Labex and Idex, ANR, R{\'e}gion Auvergne and Fondation Partager le Savoir, France; DFG and AvH Foundation, Germany; Herakleitos, Thales and Aristeia programmes co-financed by EU-ESF and the Greek NSRF; BSF, GIF and Minerva, Israel; BRF, Norway; CERCA Programme Generalitat de Catalunya, Generalitat Valenciana, Spain; the Royal Society and Leverhulme Trust, United Kingdom.

The crucial computing support from all WLCG partners is acknowledged gratefully, in particular from CERN, the ATLAS Tier-1 facilities at TRIUMF (Canada), NDGF (Denmark, Norway, Sweden), CC-IN2P3 (France), KIT/GridKA (Germany), INFN-CNAF (Italy), NL-T1 (Netherlands), PIC (Spain), ASGC (Taiwan), RAL (UK) and BNL (USA), the Tier-2 facilities worldwide and large non-WLCG resource providers. Major contributors of computing resources are listed in Ref.~\cite{ATL-GEN-PUB-2016-002}.

\printbibliography

\clearpage

\begin{flushleft}
{\Large The ATLAS Collaboration}

\bigskip

M.~Aaboud$^\textrm{\scriptsize 34d}$,
G.~Aad$^\textrm{\scriptsize 99}$,
B.~Abbott$^\textrm{\scriptsize 124}$,
O.~Abdinov$^\textrm{\scriptsize 13,*}$,
B.~Abeloos$^\textrm{\scriptsize 128}$,
D.K.~Abhayasinghe$^\textrm{\scriptsize 91}$,
S.H.~Abidi$^\textrm{\scriptsize 164}$,
O.S.~AbouZeid$^\textrm{\scriptsize 143}$,
N.L.~Abraham$^\textrm{\scriptsize 153}$,
H.~Abramowicz$^\textrm{\scriptsize 158}$
% ...
\bigskip
\\

% ...
$^{13}$Institute of Physics, Azerbaijan Academy of Sciences, Baku; Azerbaijan.\\
% ...
$^{34}$$^{(a)}$Facult\'e des Sciences Ain Chock, R\'eseau Universitaire de Physique des Hautes Energies - Universit\'e Hassan II, Casablanca;$^{(b)}$Centre National de l'Energie des Sciences Techniques Nucleaires, Rabat;$^{(c)}$Facult\'e des Sciences Semlalia, Universit\'e Cadi Ayyad, LPHEA-Marrakech;$^{(d)}$Facult\'e des Sciences, Universit\'e Mohamed Premier and LPTPM, Oujda;$^{(e)}$Facult\'e des sciences, Universit\'e Mohammed V, Rabat; Morocco.\\
% ...
$^{91}$Department of Physics, Royal Holloway University of London, Surrey; United Kingdom.\\
% ...
$^{99}$CPPM, Aix-Marseille Universit\'e and CNRS/IN2P3, Marseille; France.\\
% ...
$^{124}$Homer L. Dodge Department of Physics and Astronomy, University of Oklahoma, Norman OK; United States of America.\\
% ...
$^{128}$LAL, Universit\'e Paris-Sud, CNRS/IN2P3, Universit\'e Paris-Saclay, Orsay; France.\\
% ...
$^{143}$Santa Cruz Institute for Particle Physics, University of California Santa Cruz, Santa Cruz CA; United States of America.\\
% ...
$^{153}$Department of Physics and Astronomy, University of Sussex, Brighton; United Kingdom.\\
% ...
$^{158}$Raymond and Beverly Sackler School of Physics and Astronomy, Tel Aviv University, Tel Aviv; Israel.\\
% ...
$^{164}$Department of Physics, University of Toronto, Toronto ON; Canada.\\
% ...
$^{a}$ Also at Borough of Manhattan Community College, City University of New York, New York City; United States of America.\\
$^{b}$ Also at Centre for High Performance Computing, CSIR Campus, Rosebank, Cape Town; South Africa.\\
$^{c}$ Also at CERN, Geneva; Switzerland.\\
$^{d}$ Also at CPPM, Aix-Marseille Universit\'e and CNRS/IN2P3, Marseille; France.\\
$^{e}$ Also at Departament de Fisica de la Universitat Autonoma de Barcelona, Barcelona; Spain.\\
% ...
$^{*}$ Deceased

\end{flushleft}


\newpage

\end{document}